\documentclass[preprint,aps,showpacs,showkeys,nofootinbib]{revtex4}
\usepackage{epsfig}

\def\Journal#1#2#3#4{{#1} {\bf #2}, #3 (#4)}

\def\NPA{{\em Nucl.~Phys.} A}
\def\PLB{{\em Phys.~Lett.}  B}

\def\PRD{{\em Phys.~Rev.} D}
\def\PRC{{\em Phys.~Rev.} C}
\def\ZPA{{\em Z.~Phys.} A}
\def\ZPC{{\em Z.~Phys.} C}

\def\EPJC{{\em Eur.~Phys.~J.} C}

\def\JP{{\em J.~Phys.} G}

\begin{document}

\title{Statistical description with anisotropic momentum distributions
for hadron production in nucleus-nucleus collisions}

\author{B.~Schenke$^{1,2,*}$ and C.~Greiner$^{2,3,*}$}

\affiliation{$^{1}$Department of Physics, %
   University of Washington,%
   Seattle, Washington 98195-1560, %
   USA \vspace*{5mm}%
   \\
   $^{2}$Institut f\"ur Theoretische Physik, %
   Universit\"at Giessen, %
   Heinrich--Buff--Ring 16, %
   D--35392 Giessen, %
   Germany \vspace*{5mm}%
   \\
   $^{3}$Gesellschaft f\"ur Schwerionenforschung mbH, %
   Planckstr. 1, %
   D--64291 Darmstadt, %
   Germany \vspace*{10mm}}%

\begin{abstract}

The various experimental data at AGS, SPS and RHIC energies on
hadron particle yields for central heavy ion collisions are
investigated by employing a generalized statistical density
operator, that allows for a well-defined anisotropic local
momentum distribution for each particle species, specified by a
common streaming velocity parameter. The individual particle
ratios are rather insensitive to a change in this new intensive
parameter. This leads to the conclusion that the reproduction of
particle ratios by a statistical treatment does not imply the
existence of a fully isotropic local momentum distribution at
hadrochemical freeze-out, i.e. a state of almost complete thermal
equilibrium.

\end{abstract}



\maketitle


\section{Introduction and Motivation }
\label{Intro}

A principal question of the physics of (ultra-)relativistic heavy
ion collisions is whether the strong interacting hadronic, or
initially even deconfined partonic matter achieves an intermediate
or final reaction stage of almost thermal and chemical
equilibrium. One direct way to check for chemical equilibration is
to consider and analyze experimental hadronic abundances data by
fixing macroscopic intensive quantities such as temperature and
chemical potentials within a statistical model of an equilibrated
hadron gas \cite{BMS95,BMS96,BMS99,BMS01,Spieles}. The seminal
idea of employing such a statistical description to hadron
abundances goes back to Hagedorn for describing elementary
reactions and to Hagedorn and Rafelski for the situation of
relativistic heavy ion collisions \cite{HR80}. By now, numerous
results, initiated with the work of Braun-Munzinger et al
\cite{BMS95}, imply that a high degree of hadrochemical
equilibration is achieved at a so called stage of chemical
freeze-out, occurring in the central reactions of the heavy ion
experiments carried out at the AGS at Brookhaven, at the SPS at
CERN and at the RHIC at Brookhaven.

The analyses are based upon rather restrictive assumptions (for
two reviews having a critical look at the statistical model see
\cite{Rischke,Koch}): The standard statistical model~-- often
dubbed `thermal' model~-- assumes a complete chemical equilibrium
(see \cite{JR99} and references therein for allowing chemical
non-equilibrium) as well as thermal equilibrium at hadrochemical
freeze-out. The latter assumption implies the existence of
isotropic momentum distributions {\em locally} at the individual
space-time points within the fireball. The remarkable success of
such `thermal' models popularized the belief that the
applicability of such approaches also proves for having
established a state of nearly complete thermal equilibrium, i.e. a
state that has (locally) isotropic distributions in momentum
space. In principle, however, a statistical description is more
general. The statistical operator describes our knowledge about
the system rather than the system itself, making use of incomplete
information \cite{Katz}. Still, to best of our knowledge, so far
there has not been any investigation, concerning particle
abundances, that starts from possible and appropriate statistical
operators which do not simply resemble those of assumed (local)
thermal equilibrium. This states the motivation for the present
study, where we follow some general ideas of constructing
statistical distributions based on information theory \cite{Katz}.
Our picture is completely different from those that assume local
isotropic momentum distributions supplemented by some
hydrodynamical expansion profile in space-time \cite{Cleymans}.

A possibility to probe the degree of local equilibration is to
study the various stages of the reaction using microscopic
transport models \cite{Bravina,Sollfrank,BCG00} (see also for
lower bombarding energies the work \cite{Cugnon,Lang,Fuchs2}). As
it turns out, when testing `thermalization', present transport
models do not support the idea that a thermal system has been
achieved at intermediate or even at later stages of the reaction,
but show incomplete local momentum equilibration within the center
of mass frame of a small but finite central cell
\cite{Bravina,BCG00}.

To meet this insight we propose in the following to relax the
constraint of assuming \emph{locally} fully isotropic, statistical
distributions for the comparison to hadron abundances, i.e. of
assuming complete kinetic equilibrium. Certainly, at the onset of
a heavy ion reaction, the local momentum distributions of the
various particles are strongly non-isotropic. Moreover, the
various measured rapidity spectra point towards a global picture
of the fireball which shows still a considerable longitudinal
momentum excess as the spectra are much broader than those of a
pure thermal and static source. Part of this broadening could be
attributed to the buildup of a collective hydrodynamical
expansion, but the other part can in principle originate from
locally anisotropic momentum distributions. When constructing a
generalized statistical operator, we will follow this latter
possibility that the system might not have developed fully
isotropic momentum distributions locally in space-time at the
point of chemical freeze-out.


\section{Generalized statistical operator for longitudinally
\\
deformed configurations in momentum space}
\label{model}

A concept of a statistical description which incorporates at least
partially the initial longitudinal momentum excess has been
proposed by Neise \cite{Neise,Neise1} some time ago in order to
allow for a generalization of the nuclear equation of state at
finite density and temperatures for momentum anisotropic phase
space distributions. Originally, the idea of two
interpene\-trating cold fluids goes back to Lovas \cite{Lovas},
who considered both an ellipsoidally deformed Fermi sphere and two
separated Fermi spheres by introducing an additional parameter $v$
which effectively alters the momentum distribution. This latter
parametrization has been considered for various studies concerning
the effective in-medium interaction and relativistic potentials
experienced by nucleons inside a non-equilibrium environment
\cite{Wolter}. For relativistic heavy ion collisions at GSI
energies, i.e. for Au+Au--collisions at 1 AGeV, the relative
velocity $v_{rel}$ has been analyzed within a microscopic
transport model and was found to be $\approx 0.6 \, c$ at the
stage of maximum compression \cite{Fuchs2}, manifesting a full
nonequilibrium situation at intermediate reaction times.

The main idea in \cite{Neise} is to
interpret the velocity parameter as an additional statistical
Lagrange multiplier $\nu $ by
introducing a generalized grand canonical potential
\begin{equation}
\label{Omgen}
\Omega \, = \, E \, - \, TS \, - \, \mu N \, - \nu O \, \, \, ,
\end{equation}
where $O$ represents the (thermodynamic) expectation value of an
observable $O$, which the system should assume within the grand
canonical average as an additional constraint. With this
statistical approach various selfconsistency relations are
automatically fulfilled. If the observable $O$ is given by a
one-particle operator, $O = \sum_i n_i O_i $, minimization of
(\ref{Omgen}) for a gas of independent hadronic particles with a
given set of one-particle energy states $\epsilon _i $ results in
the generalized distribution \cite{Neise,Neise1,Weidlich}
\begin{equation}
\label{Omgendist}
\frac{\partial \Omega }{\partial n_i } = 0 \quad \Rightarrow \quad
n_{i} \, = \, \frac{1}{
\exp \left[ \beta (\epsilon _i   - \mu - \nu O_i\right] \pm 1}
\, \, \, ,
\end{equation}
depending on the quantum statistics of the considered hadronic
particle. The formalism can be easily generalized to allow for
self-consitent mean-field type interactions
\cite{Neise,Neise1,Wolter}. To allow for a specified momentum
anisotropy, one can choose an adequate one-particle operator $O$
characterizing the momentum anisotropy by taking the difference
between the mean value of the momenta with a positive and negative
longitudinal ($z$) component \cite{Neise} respectively, i.e.
\begin{equation}
\label{Operator}
O[p]\, =\,
\sum_p \left[ p_z \theta (p_z) - p_z \theta (-p_z) \right] a_p^+ a_p
\, = \, \sum_p |p_z| a_p^+ a_p \, \, \, .
\end{equation}
According to (\ref{Omgendist}) one ends up with the generalized
momentum distribution
\begin{equation}
\label{gendistr}
n_{\bf p} \, = \, \frac{1}{
\exp \left[ \beta (\epsilon _p  - v |p_z| - \mu ) \right] \pm 1} \, \, \, ,
\end{equation}
with the conjugated intensive quantity $\nu \equiv v$.

\begin{figure}[htb]
\includegraphics[width=60mm]{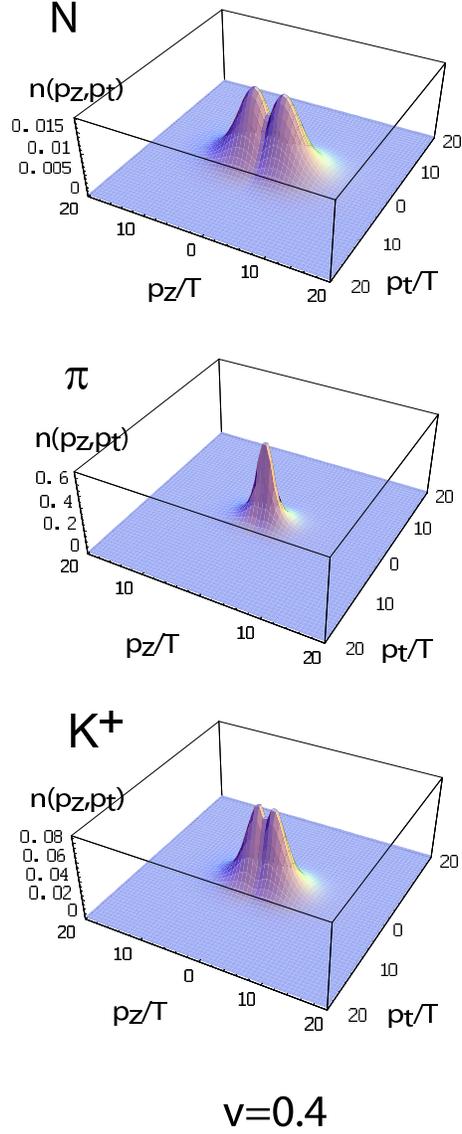}
\caption{Generalized statistical momentum space distribution
with $v=0.4$ for the nucleons, the pions and the $K^+$.
Temperature parameter $\beta_0 $ and chemical potentials
are taken as those extracted for $v=0$ being typical for Pb+Pb collision
at SPS energies
\protect\cite{BMS99}.}
\label{fig_aniso}
\end{figure}

A more appropriate relativistic notation for the distribution can
be obtained by the following more intuitive and explicit
construction of the same density operator \cite{Neise1}. Suppose
that the homogeneous medium consists of two subsystems in momentum
space (dubbed `projectile' and `target' component indicating their
origin), which are specified by their collective flow velocity
$v^{(P)\mu} \, = \, \gamma (1,0,0,v)$ and $v^{(T) \mu }\, = \,
\gamma (1,0,0,-v)$ in the combined center of mass frame with
$\gamma =(1-v^2)^{-1/2} $. It becomes obvious that the intensive
parameter $v$ corresponds to the velocity of each subsystem. In
the special c.m. frame only the positive z-component, respectively
negative z-component, of the (assumed) single-particle momenta
contribute to the projectile subsystem, respectively target
subsystem, i.e. both subsystems are separated by the plane
$p_z=0$. For the grand canonical potential one takes the sum of
the two components, i.e.
\begin{equation}
\label{GCpot2}
\Omega \, = \, v^{(P)\nu } P_\nu ^{(P)} - T_0S^{(P)}  - \mu_0 N_B^{(P)}
+ \, v^{(T)\nu } P_\nu ^{(T)} - T_0S^{(T)}  - \mu_0 N_B^{(T)}
\, \, \, ,
\end{equation}
from which the statistical density operator immediately results
as
\begin{equation}
\label{densm}
\rho  \, = \, \frac{1}{{\bf Z} }
\exp \left\{ - \beta_0 \left(
v^{(P)\nu } P_\nu ^{(P)} + v^{(T)\nu } P_\nu ^{(T)} \right)
+\beta_0 \mu_0 N_B \right\} \, \, \, .
\end{equation}
For a noninteracting resonance gas of
independent hadronic particles
one has for the total momentum
\begin{equation}
\label{PGes}
P_\mu ^{(P/T)} \, = \, \sum_h \sum _{\bf p} \, p_{\mu }^{h} \,  n_{\bf p}^{h}
\, \theta ( \small{ +/-}\,  p_z) \, \, \, .
\end{equation}
A similar expression holds for the entropy and baryon number.
For the grand canonical potential (\ref{GCpot2})
the contributions for the entropy and the baryon number are entering simply
additively, and, as $\theta (p_z) + \theta(-p_z) =1$,
the explicit
decomposition is only relevant for the sum
$v^{(P)\nu } P_\nu ^{(P)}+ v^{(T)\nu } P_\nu ^{(T)}$.
Putting everything together, from
(\ref{densm}) one ends up with the expression
\begin{equation}
\label{gendistra}
n_{\bf p}^{h} \, = \, \frac{1}{
\exp \left\{ \beta _0 \gamma (\epsilon _p^{h}  - v |p_z|)  - \beta_0 \mu_0
\right\} \pm 1}
\end{equation}
for the individual momentum distributions. (\ref{gendistra}) is
completely equivalent to (\ref{gendistr}): $\beta_0 $ and $\mu _0
$ represent the true inverse `temperature' and (baryo-)chemical
potential defined in the local rest frame of each component,
whereas $\beta $ and $\mu $ in (\ref{gendistr}) are effective
parameters modified by $\gamma $. A straightforward consideration
shows that the momentum distribution becomes elongated in the
longitudinal direction by the Lorenz factor $\gamma $. As an
illustration, fig.~\ref{fig_aniso} depicts a situation typical for
central Pb+Pb collisions at SPS energies for the case of nucleons,
pions and kaons with an intensive parameter chosen to be $v=0.4$
c. The heavier the hadronic particle, the more pronounced the two
peaks $\pm m^{h}\gamma v $ in the distribution of the `two fluids'
manifest.

Summarizing, the velocity $v$ should be thought as the intrinsic
(Lagrange-)parameter for characterizing any potential
non-equilibrium situation in local longitudinal momentum space
within the statistical description. We consider it as the variable
for characterizing a possible intermediate situation of the
momentum distributions at hadrocemical freeze-out in the local
comoving rest frame of a moving spatial cell within the spatially
expanding fireball. This cell is then considered as a
characteristic representative of the whole system, like for
example within a boost-invariant Bjorken geometry: A
boost-invariant but locally momentum anisotropic distribution is
given by the standard substitution
\begin{equation}
 n^h_{\bf p}=n^h(p_\perp,p_z)=n^h(p_\perp,y)\rightarrow
 f^h(p_\perp,y;z,t)=n^h(p_\perp,y-\eta)
 \label{subs}
\end{equation}
of Lorentz-boosted comoving cells, and assuming the intrinsic
parameters $T,\mu_B$ and $v$ only to depend on the eigentime
$\tau$.

The parameter $\gamma v$ is certainly different and considerably
smaller than in the very initial situation of the heavy ion
reaction due to the (unknown) amount of dissipation occurring in
the (violent) initial stages. The mismatch compared to a fully
momentum isotropic situation should relax continuously in time due
to collisions among the various particles, although the system
might never actually become locally completely isotropic
\cite{Bravina,BCG00}, as outlined in the introduction. In the
following we thus will consider $v$ as a free parameter.

Finally it is worthwhile to point out the strict meaning of a
thermodynamical equilibrium description compared to statistical
descriptions in general \cite{Katz}. A statistical operator,
relying on the concept of incomplete information, assumes the
existence of some average constraints $\langle O_i\rangle$. The
density operator then has the standard form $\rho =\exp (\Omega -
\sum_i \nu _i O_i)$ \cite{Katz}. Generally, the operator $O_i$
does not need to commute with the Hamiltonian $H$. Being more
restrictive, one has a stationary ensemble when taking only
constants of motion for the $O_i$; in other words, the Liouville
equation for the density matrix
\begin{equation}
\label{Liouville}
\frac{d \rho}{dt}=\{H,\rho\}=0 \, \, \, ,
\end{equation}
has to be fulfilled for a true thermodynamical equilibrium
situation. This is not the case for the chosen momentum anisotropy
operator $O$ reflecting the fact that the momentum anisotropy
defined might actually be a time dependent quantity. As outlined
above, collisions among the particles in the (strongly)
interacting matter are ultimately destroying the stationarity. A
requirement for stationarity is fulfilled when considering
spatially homogeneous matter and employing approximate effective
one-particle (and thus collision-less) Hamiltonians of mean-field
type or a simple quasi-free Hamiltonian for describing the hadron
resonance gas. On the other hand, the statistical assumption is
made when postulating the existence and minimization of the
generalized grand canonical potential (\ref{Omgen}) with the
constraint that locally a finite mismatch in momenta survives at
hadro-chemical freeze-out.


\section{Influence of the velocity parameter on particle ratios}

Following the bootstrap idea of incorporating the strong
interactions among the constituents (see eg \cite{Weidlich}), we
assume a `noninteracting' hadron (Hagedorn-like) resonance gas for
describing the excited hot hadronic matter inside the fireball
\cite{BMS95,BMS96,BMS99,BMS01,Spieles}. The chemical freeze-out of
particles is assumed to happen for a fixed temperature $T$. For
the sake of simplicity we do not intend to distinguish between a
$4\pi $-analysis or an analysis of particle ratios being measured
only close to midrapidity: An approximate Bjorken geometry as
given in (\ref{subs}) would give the same answer for both
analyses.

We choose the description of an isospin symmetric hadronic gas and
include baryonic and mesonic resonances of masses up to 2 GeV
\cite{Spieles,GS91}. For the various particle number densities
described by the generalized momentum distributions
(\ref{gendistra}) one has
\begin{equation}
    \rho_i \, = \, \frac{g_i}{(2 \pi)^3}\int d^3p \frac{1}{\exp(\beta _0
    \gamma (E_i-v|p_z|)-\beta _0 \mu_{i; 0})\pm 1} \, ,
\label{gendens}
\end{equation}
with spin degeneracy $g_i$,
$\beta _0 =\frac{1}{k_BT}$ and $E_i=\sqrt{p_\perp^2+p_z^2+m_i^2}$.
The chemical potential of each hadronic particle
is specified by its baryon number content $N_B$ and its strangeness
number content $N_S$ via $\mu_{i; 0}= N_B \mu_B + N_S \mu_S $,
assuming isospin symmetric hadronic matter ($\mu_I =0$).

Changing the velocity parameter, the various particle number
densities do increase with increasing $v$ to almost $2 \gamma $ for
velocities close to $c$.
The total intrinsic energy density, on the other hand, will scale
with almost $2\gamma ^2$ for such large velocities.

We are now in position to check for the sensitivity in extracting
the thermodynamical intensive parameters like the `temperature'
parameter $\beta_0 ^{-1}$ and the chemical potentials when varying
the velocity parameter $v$. It is worthwhile to first discuss two
limiting situations: For $v=0$ the standard statistical model
description results. For velocity parameters being sufficiently
large, the two interpenetrating components fully separate in
momentum space, so that the hadronic particle ratios will not be
altered when employing the same temperature $\beta_0 ^{-1}$ and
chemical potentials as for $v=0$. Each of the separate subsystems
then describes the situation of a Lorentz boosted momentum
isotropic fireball, i.e. the situation for $v=0$. The question
remains for the true intermediate situations, where one cannot
really distinguish among two separate components. As a further
limiting case, employing a classical Maxwell-Boltzmann
approximation,
$$
n_{\bf p} \, \approx \,
\exp \left(- \beta _0 \gamma (E_i  - v |p_z|)\right)
\, \exp (  \beta_0 \mu_{i,0} ) \, \, \, ,
$$
number ratios of particle and their anti-particles are not
altered by a change in the velocity parameter, being
solely dependent on the exponential factor
containing the chemical potential.
We therefore expect only some smaller  kinematical
sensitivity for particle
ratios with unequal masses.

\begin{figure}[htb]
\includegraphics[width=130mm]{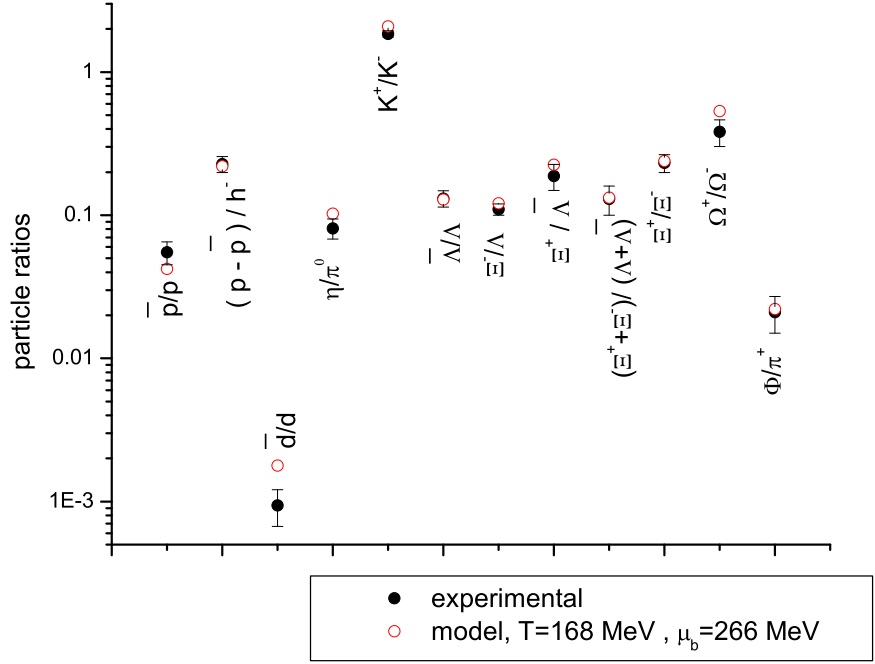}
\caption{Characteristic particle ratios for
Pb+Pb collisions at SPS energies
obtained with velocity parameter $v=0$.
Temperature parameter $\beta_0 $ and chemical potentials
are taken the same as in
\protect\cite{BMS99}.}
\label{fig_ratio}
\end{figure}

\begin{figure}[t]
\includegraphics[width=100mm]{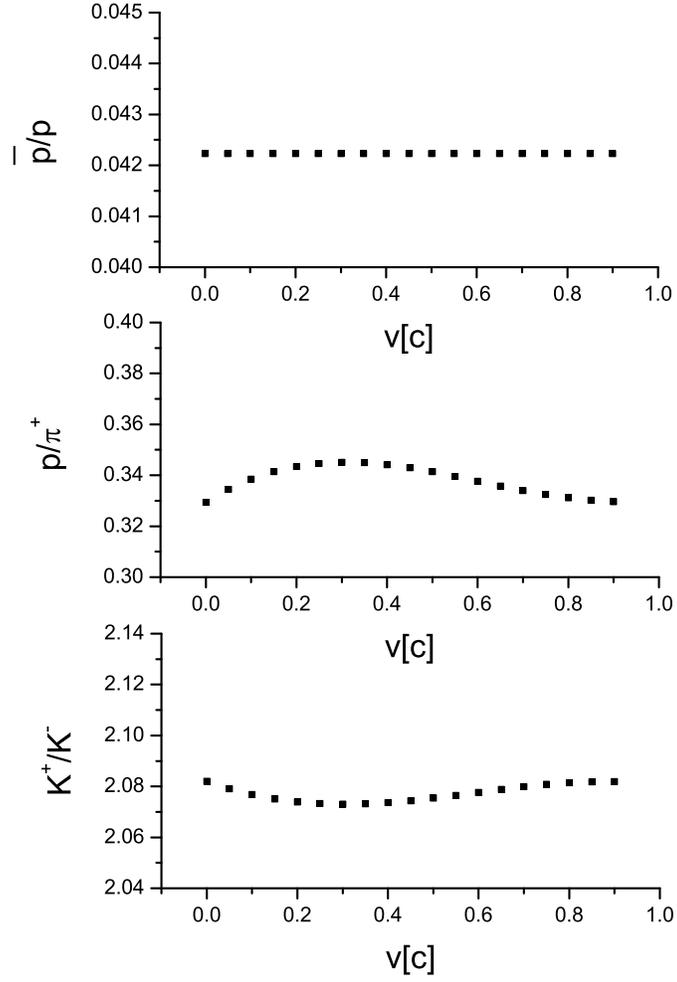}
\caption{(Weak) Dependence of three typical particle ratios
($\bar{p}/p; \, p/\pi ^+ ; \, K^+/ K^- $)
varying the velocity parameter from $v=0 \, c$ to $v=0.9 \, c$.
Temperature parameter and chemical potentials
are taken the same as those of fig.~\ref{fig_ratio}
with $v=0$.}
\label{fig_rat}
\end{figure}

For testing the potential influence of an anisotropic momentum
distribution on the hadronic particle ratios the strategy is as
follows: We first employ the known values for the temperature
$T_0$ and potential $\mu_{B,0}$ of \cite{BMS95,BMS96,BMS99,BMS01}
within the present description taking  $v=0$ for describing the
various situations of heavy ion experiments carried out at the
AGS, at the SPS and at RHIC. As one particular example we show in
fig.~\ref{fig_ratio} characteristic particle ratios in comparison
to the experimental results for central Pb+Pb collisions at SPS
energies. We find good agreement to the various hadronic particle
ratios (in quality like those given in
\cite{BMS95,BMS96,BMS99,BMS01}) and we mention the well-known fact
that feeding from all relevant hadrons and resonances to the
abundances of `stable' (i.e. detectable) particles is of crucial
importance. With the temperature and the chemical potentials held
fixed, we now change the velocity parameter in order to see the
sensitivity on the particle ratios. In the following we
concentrate on the $\bar{p}/p$~-- , the $p/\pi^+$~-- and the $K^+/
K^-$~--ratio. The first two ratios typically fix the temperature
and the baryochemical potential. The last fixes the strange
chemical potential, which, however, enters as a constraint as the
system carries no overall net strangeness. In fig.~\ref{fig_rat}
the actual (non-)dependence of these exemplary particle ratios on
varying the velocity parameter is depicted. As expected from the
above estimates, the largest deviation, about 6\%, one finds for
the $p/\pi^+$ ratio (for the situation of a Pb-Pb collisions at
SPS). We have tested for all the various other particle ratios
shown in fig.~\ref{fig_ratio} and find, if at all, only a marginal
dependence on the velocity parameter $v$. $p/\pi^+$ shows the
`strongest' sensitivity. The same situation is met when looking at
ratios measured at AGS or at RHIC. Most of the other ratios, such
as the shown $\bar{p}/p$ and $K^+/K^-$, only vary by about 1\% or
even less. In tab.~\ref{tab:ratios} the above three particle
ratios as obtained within our investigation are summarized. There
is no remarkable change of the particle ratios at any of the
considered collision energies.

Hence, the quality of a possible detailed fit to the particle
ratios including the velocity as a free parameter would not be
changed: An extraction of $v$ would be completely ambiguous.


\section{Summary and Conclusions}
\label{conclusion}

We have constructed a statistical description of a hadronic
resonance gas allowing for a finite longitudinal momentum excess
compared to commonly employed isotropic (`thermal') distributions.
This is achieved via a velocity parameter acting as a common
Lagrange multiplier for fixing a momentum anisotropy for the local
momentum distributions . As microscopic transport models do not
fully support the idea of a complete local momentum equilibration
at any stage of a central heavy ion reaction, the presented
approach relaxes this typical constraint of a standard statistical
model. From a physics point of view one can argue that the amount
of dissipation or collisions is simply not enough to achieve full
momentum equilibrium locally in space. Going further with the
argument, there exists always the principal question when applying
a statistical treatment for hadron production whether it
represents a reflection of a true dynamical and stationary
equilibrium state or, more generally, whether it is only a
reflection of a state minimizing the content of information (i.e.
maximizing the `entropy') within a set of appropriate, global
constraints \cite{Katz}, as also employed in this study. This is
particularly obvious when applying such concepts to $e^+ e^-$--,
$pp$-- or $p\bar{p}$--collisions, as proposed seminally by
Hagedorn and described recently by Becattini \cite{Be96} within a
micro-canonical treatment. For such `systems' one can not expect
the occurrence of a true dynamical equilibrium state (fulfilling
approximately (\ref{Liouville})) at any stage of the reaction.

It turns out that the hadronic particle composition is more or
less completely insensitive to varying the velocity parameter from
zero to nearly speed of light. The extraction of this parameter
can not be achieved by looking at particle number ratios. On the
other hand, the extraction of the `temperature' parameter as well
as the (baryo-)chemical potential, however, are the same as with
the standard approach. They are thus the significant and
appropriate statistical quantities for dealing with the various
number ratios characterizing the intermediate stage of the
reaction at hadrochemical freeze-out.

The popular conclusion that the many successes in describing
particle ratios within a `thermal' model give evidence for the
existence of a state of thermal equilibrium at hadrochemical
freeze-out is, however, not true. Local isotropic momentum
distributions reflect merely a simplifying assumption of the
applied statistical operator. With this observation in mind, the
standard `extraction' of total particle number densities or energy
densities, respectively, in absolute terms achieved inside the
hadronic fireball at chemical freeze-out becomes also unsteady.
Our study underlines the necessity of understanding relativistic
heavy ion collision experiments by means of detailed microscopic
transport models.


\subsection*{Acknowledgments}

This work has been partially supported by BMBF, by the
European Graduate School Copenhagen/Giessen `Complex Systems
of Hadrons and Nuclei' and by GSI.
The authors thank U.~Mosel for discussions and his continuous interest
in this work.
CG acknowledges discussions with C.~Fuchs, J.~Knoll, M.~Lutz, H.~St\"ocker
and R.~Stock.



\newpage
\vspace*{20mm}

\begin{table}[htb]
\begin{center}
\begin{tabular}{|c|c|c|c|}
        \hline
                       & model for $v=0$    & model for parameter $v$ &
 experiment          \\

 & & with largest deviation in \% & \\
\hline \hline
{\bf AGS}& \multicolumn{3}{c|}{$T=120$ MeV, $\mu_B=540$ MeV, $\mu_S=108 $
 MeV}
\\ \hline
           $\bar{p}/p$      & 0.000124
& 0.04652     &  $0.00045 \pm 0.00005 $    \\
            $p/\pi^+$          & 1.0535
& 5.9178  &  $ $                       \\
            $K^+/K^-$        & 5.5362
& -0.5315     &  $4.4 \pm 0.4 $            \\
        \hline \hline
{\bf SPS I}& \multicolumn{3}{c|}{$T=160$ MeV, $\mu_B=170$ MeV, $\mu_S=38 $ MeV}
\\ \hline
            $\bar{p}/p$      & 0.1195
& 0.0048     &  $0.12 \pm 0.02 $       \\
            $p/\pi^+$          & 0.1863
& 6.1940    &  $0.18 \pm 0.03 $       \\
            $K^+/K^-$        & 1.51
& -0.2680   &  $1.67 \pm 0.15 $       \\
        \hline \hline
{\bf SPS II}& \multicolumn{3}{c|}{$T=168$ MeV, $\mu_B=266$ MeV, $\mu_S=71.1 $
 MeV}
\\ \hline
          $\bar{p}/p$      & 0.0422
& 0.0104     &  $0.055 \pm 0.01 $    \\
            $p/\pi^+$          & 0.3294
& 4.7817     &  $ $                  \\
            $K^+/K^-$        & 2.0820
& -0.4319     &  $1.85 \pm 0.09 $     \\
        \hline \hline

{\bf RHIC}& \multicolumn{3}{c|}{$T=174$ MeV, $\mu_B=46$ MeV, $\mu_S=13,1 $
 MeV}
\\ \hline
            $\bar{p}/p$      & 0.5845
& 0.8568     &  $0.60 \pm 0.07 $          \\
            $p/\pi^+$          & 0.1129
& 5.4299     &  $ $                       \\
            $K^+/K^-$        & 1.1282
& -0.1297     &  $1.136 \pm 0.070 $        \\
        \hline
\end{tabular}
\vspace*{10mm}
       \caption{Characteristic particle ratios from
various heavy collision experiments (from top to bottom):
(1) $Si+Au$ at the AGS, parameters from \protect\cite{BMS95};
(2) $S+Au$ at the SPS, parameters from \protect\cite{BMS96};
(3) $Pb+Pb$ at the SPS, parameters from \protect\cite{BMS99};
(4) $Pb+Pb$ at RHIC, parameters from \protect\cite{BMS01}.
The maximal deviation (in percent) by changing the velocity parameter
is indicated.}
\label{tab:ratios}
\end{center}
\end{table}


\newpage
\vspace*{20mm}

\newpage
\vspace*{20mm}

\newpage
\vspace*{20mm}

\end{document}